\documentclass{article}
\usepackage[utf8]{inputenc}
\usepackage{booktabs}
\usepackage{graphicx}
\usepackage{hyperref}
\usepackage{amsmath}

\newtheorem{assumption}{Assumption}

\graphicspath{ {./figures/} }

\newcommand{\zi}[1]{\textit{#1}}

\newcommand{\wx}[1]{\textsf{#1}}
\newcommand{\REfull}{Requirements Engineering}
\newcommand{\RE}{Requirements Engineering}
\newcommand{\RPfull}{Requirements Problem}
\newcommand{\RP}{Requirements Problem}
\newcommand{\ZJ}{ZJ}
\newcommand{\ZJRPfull}{Default \RPfull}
\newcommand{\ZJRP}{Default \RPfull}
\newcommand{\RC}{Requirements Contract}

\title{Requirements Contracts: Definition, Design, and Analysis}
\author{Ivan J. Jureta\\
Fonds de la Recherche Scientifique -- FNRS, Brussels, Belgium,\\
Universit\'{e} de Namur, Belgium,\\
STEMCELL Technologies Inc., Vancouver, Canada\\
\texttt{ivan@ivanjureta.com}, \url{http://ivanjureta.com}}

\date{\today}

\begin{document}

\maketitle

\begin{abstract}
What are the necessary and sufficient conditions for a proposition to be called a requirement? In Requirements Engineering research, a proposition is a requirement if and only if specific grammatical and/or communication conditions hold. I offer an alternative, that a proposition is a requirement if and only if specific contractual, economic, and engineering relationships hold. I introduce and define the concept of ''Requirements Contract'' which defines these conditions. I argue that seeing requirements as propositions governed by specific types of contracts leads to new and interesting questions for the field, and relates requirements engineering to such topics as economic incentives, interest alignment, principal agent problem, and decision-making with incomplete information.
% \keywords{Requirements \and Economics \and Law \and Decision-Making.}
\end{abstract}

\tableofcontents

\section{Background}\label{s:background}
In research on Requirements Engineering, sentences such as ``the system should do A'', ``the software should do B'', or ``the system must never do C'' are called requirements. This is an example:
\begin{quote}
A meeting scheduler should know the constraints of the various participants invited to the meeting within some deadline $d$ after invitation. \cite{van1998managing}
\end{quote}

Requirements Engineering, as a research discipline, came out of the need to be clear and rigorous about how to define, document, change requirements for software and hardware, and how to evaluate if and how much requirements are satisfied. But requirements are not confined to software and hardware. Law defines requirements on legal entities, governments on the behaviour of citizens, schools on the behaviour of students, and so on. 

When a proposition is called a requirement, it conveys what someone expects, needs, wants, asks for. 

This general idea that requirements set expectations about the thing to be made or bought, and/or of behaviour or competence to exhibit\footnote{Consider, for example, the following passage on Canadian citizenship requirements (see \url{https://www.cic.gc.ca/}: ``To become a Canadian citizen, you must: be a permanent resident, have lived in Canada for 3 out of the last 5 years, have filed your taxes (if you need to), pass a test on your rights, responsibilities and knowledge of Canada, prove your language skills''. Those requirements are not about things to make or buy, but on competences and behaviours.} has been around ever since the early research publications in the field. Ross and Schoman wrote ``requirements definition is a careful assessment of the needs that a system is to fulfil. It must say why a system is needed, based on current or foreseen conditions'' \cite{ross1977structured}. Brooks claimed that ``[the] hardest single part of building a software system is deciding precisely what to build'' \cite{brooks1987}. Zave wrote that ``Requirements engineering is the branch of software engineering concerned with the real-world goals for, functions of, and constraints on software systems'' \cite{zave1995classification} Hsia, Davis, and Kung wrote that ``Requirements engineering is the disciplined application of proven principles, methods, tools and notations to describe a proposed system's intended behavior and its associated constraints'' \cite{hsia1993status} Continuing Ross and Schoman's line of thought, in arguing why it matters to think about goals of a system-to-be, van Lamsweerde claimed that ``goals provide the rationale for requirements [...] a requirement appears because of some underlying goal which provides a base for it'' and he saw goals as that which relates the system to the organization which uses it \cite{van2001goal}. Sommerville argued that ``before developing any system, you must understand what the system is supposed to do and how its use can support the goals of the individuals or business that will pay for that system'' \cite{sommerville2005integrated}.

\section{Problem}\label{s:problem}
We can have an arbitrary number of propositions that refer to what is needed, wanted, desired. Such proposition are common; they are there when people complain, for example, as well as when they set goals to achieve, or rules for how something should or shouldn't be done. 

This abundance of requirements-like propositions begs the following question: \textit{Are all these propositions always requirements?} Or you can look at the same problem the other way: \textit{How does a proposition become a requirement?} These questions are the focus of this paper.

If \textit{grammatical mood} fully determines whether a proposition has the role of requirement, then the only condition that should be true for a proposition $p$ to get that role, is that $p$ is communicated in the right grammatical mood. This is what Michael Jackson argued in several papers in the 1990s \cite{jackson1995problems,jackson1997meaning}, including in the influential ``Four dark corners'' with Pamela Zave \cite{zave1997four}, and later, ``A Reference Model for Requirements and Specifications'' \cite{gunter2000reference} with Carl Gunter, Elsa Gunter, and Pamela Zave. 

If the \textit{type of speech act} determines if a communicated proposition gets the role of requirement, then all that's needed for $p$ to get that role, is that the right speech act is used. John Mylopoulos and I supported this position with various co-authors in our publications on the core ontology for requirements engineering \cite{jureta2008revisiting,jureta2009core} and on the \textit{Techne} family of requirements modelling languages \cite{jureta2010techne,jureta2014requirements}.

If the assignment of a goal to a software agent is enough for a the goal to become a requirement, then as long as the goal is refined to such extent that it can be satisfied by functionality implemented by software, this is enough for it to be called a requirement. That  is Axel van Lamsweerde's position -- ``[a] goal under responsibility of a single agent in the software-to-be becomes a requirement whereas a goal under responsibility of a single agent in the environment of the software-to-be becomes an assumption'' \cite{van2001goal}. A goal for van Lamsweerde is exactly what Jackson called requirements (see above): ``A goal is an objective the system under consideration should achieve. Goal formulations thus refer to intended properties to be ensured; they are optative statements as opposed to indicative ones, and bounded by the subject matter'' \cite{van2001goal}.

Other than minor differences\footnote{The only one that stands out is van Lamsweerde's constraint to be able to assign a goal to a software agent, which is essentially about level of detail, i.e., he asks to have the goal refined up to a threshold level of detail tied to the feasibility of software to implement functionality that supports the original goal (see, e.g., \cite{darimont1996formal}). This makes no difference, because he still sees goals the same way Jackson sees requirements.}, all three positions revolve around the same idea: a proposition is called a requirement, if it is such that we understand it as referring to something that is needed, desired, or otherwise asked for; that seems to be both the necessary and sufficient condition for a proposition to become a requirement.

What is wrong with the above? In simplest terms, you can want and say that you want $p$ to be satisfied, yet no one may need to invest effort to do so. This ties closely with such intuitively appealing and simple observations as that we cannot satisfy all needs we may think of. As the Rolling Stones say, you can't always get what you want.

More specifically, the problem is that the position above -- that all we need for a proposition to become a requirement is that it is in the right grammatical mood or speech act -- clashes with the following common sense observation.

\begin{quote}
\textit{A proposition $p$, which may refer to something I would like to see satisfied in the future, cannot be a requirement for you to satisfy just because I wish this to be so and I say that I wish it to be so.}
\end{quote}

Grammar and speech acts alone are not enough for someone to satisfy that which another person said they want, need, or desire. Something more is needed, something that will lead others to want to invest effort to satisfy these expressed goals, needs, or whatever we want to call them. It follows that propositions become requirements for some other reasons than simply because someone wants them to be satisfied. 

The problem, then, is this: \textit{What are the necessary and sufficient conditions for a proposition to become a requirement?}

\section{Solution Outline}
The following paragraph summarises what is missing when thinking about requirements as propositions communicated to convey needs.

\begin{quote}
\textit{For a proposition $p$ to become my requirement for you, I have to have been given specific rights, which I can then exercise to give $p$ the role of requirement, and you accept the obligation to bring about $p$. Both you and I should get value out of this arrangement: I will if my requirement is satisfied, and you will if there is something you get out of it, i.e., there is some transfer of value to you for satisfying that requirement. Finally, you should be confident enough that you will be able to satisfy my requirement through effort that you can invest, given the value that you expect.}
\end{quote}

To take the above seriously is to accept, as I will argue, that a proposition $p$ gets the role of requirement within a nexus of contractual, economic, and engineering relationships. There needs to be a contract which distributes rights and responsibilities, and makes economic relationships binding for the parties involved; there need to be economic relationships, that is, expectations of getting value for anyone to want to enter the contract; and there have to be, broadly speaking, engineering relationships satisfied, between the understanding of what requirements are about, and assumptions of what should be designed, built, and eventually used or run to satisfy them.

As a result, a requirement is not just a proposition denoting desired future conditions, a proposition that is the content, so to speak, of what someone may wish. We cannot say that a proposition \textit{is} a requirement at all. The proposition remains a proposition, but can be given the role of requirement by those who enact roles defined in a contract, exercise their rights and act according to their obligations. A proposition maintains the role of requirement as long as the contract remains applicable. Requirements, in short, come out of contracts, not wishes.

\section{Paper Outline}
Why should you take the solution outline above seriously? Why do we need contracts when focusing on Requirements \textit{Engineering}? Can't we deal with the engineering of requirements while ignoring the contracting around their satisfaction? Section \ref{s:rationale:contract} focuses on these questions, and develops the argument that the engineering relationships for a proposition to have the role of requirement are inseparable from the contractual ones.

If the engineering and contractual relationships are inseparable, then can't we at least keep \textit{economic} relationships out of the picture? Can we abstract those away and deal with them separately? Section \ref{s:rationale:economics} is dedicated to these questions, and concludes that contractual and economic relationships are inseparable. Therefore, if we take engineering and contractual relationships, then we need to take economic ones along as well.

Section \ref{s:solution} gives the necessary and sufficient conditions for a proposition to get the role of requirement. 

An important implication of seeing requirements as roles resulting from a nexus of contractual, economic, and engineering relationships, is that the so-called ``requirements problem'' looks like a limited treatment of a more complicated problem. I will argue that, unfortunately, we cannot be defining and solving that engineering problem without defining and solving the contractual and economic ones with it. One of the interesting, and I believe constructive consequences is that we need to have a broader and richer discussion in Requirements Engineering research, one which does not set aside economic and contractual relationships and constraints that affect so many parameters of how the engineering problem gets formulated and solved. 
% Section \ref{s:design} discusses the design of that nexus of relationships that yield requirements.

The paper closes with a discussion of the weaknesses of the proposal in this paper, and mentions open questions that I encourage colleagues to consider.

\section{Rationale}\label{s:rationale}
To argue that requirements cannot exist without contractual and economic relationships, we take a step back and recall the basic ideas in \RE{} research, on what the central problem is, i.e., what you are up against when doing \RE. 

\subsection{Background to the Rationale}\label{s:rationale:background}
\REfull{} focuses on eliciting, modelling, and analysing the requirements and environment of a system-to-be in order to design its specification. 

It is on the basis of its specification that the system is built, updated, changed, its new releases planned, made, announced, rolled out. Specifications can take different forms, ranging from minimalist to-do lists that hint at expectations and subsume implicit engineering solutions, to elaborately structured documentation on responsibilities of positions in the value chain, guidelines for employee coordination and collaboration, as well as formal software specifications made for use with a model checker. 

The design of the specification, usually called the \RPfull{}, is a complex problem solving task, as it involves, for each new system-to-be, the discovery and exploration of, and decision making in, new and ill-defined problem and solution spaces. 

Difficulties involved in solving an \RP{} instance are illustrated by the variety of topics studied in \RE{} research, such as requirements elicitation \cite{goguen1993techniques, hickey2004unified, davis2006effectiveness}, categorization \cite{dardenne1993goal, zave1997four, jureta2008revisiting}, vagueness and ambiguity \cite{mylopoulos1992representing, letier2004reasoning, jureta2007clarifying}, prioritization \cite{karlsson1998evaluation, berander2005requirements, herrmann2008requirements}, negotiation \cite{leite1991requirements, boehm1995software, jureta2009analysis}, responsibility allocation \cite{dardenne1993goal, castro2002towards, fuxman2004specifying}, cost estimation \cite{boehm1984software, boehm2000software, sindre2005eliciting}, conflicts and inconsistency \cite{nuseibeh1994framework, heitmeyer1996automated, van1998managing}, comparison \cite{mylopoulos1992representing, letier2004reasoning, liaskos2010integrating}, satisfaction evaluation \cite{boehm1976quantitative, mylopoulos1992representing, krogstie1995towards}, operationalization \cite{giorgini2003reasoning, fuxman2004specifying, ernst2013agile}, traceability \cite{gotel1994analysis, ramesh2001toward, cleland2005utilizing}, and change \cite{cheng2009software, whittle2009relax, brun2009engineering}.

The \textit{de facto} default view in \RE, is that the specification of the solution to build or buy, is produced \textit{incrementally}, starting from a limited set of incomplete, inconsistent, and imprecise information about the requirements and the system's operating environment, and that each design step reduces incompleteness, removes inconsistencies, and improves precision, towards the specification of the system \cite{boehm1988spiral, dardenne1993goal, greenspan1994formal, nuseibeh1994framework, finkelstein1994inconsistency, zave1997four, van2001goal, castro2002towards, robinson2003requirements, jureta2010techne, ernst2013agile}.

This important and general conceptualisation of the aim in \RE{} is most clearly formulated in Zave \& Jackson's ``Four dark corners of requirements engineering'' \cite{zave1997four} mentioned in Section \ref{s:problem}. Their view, denoted \ZJ{} hereafter, is echoed in some of the most influential research in the field, which both preceded and followed the said paper, including, for example, contributions from Boehm et al. \cite{boehm1988spiral,boehm1995software}, van Lamsweerde et al. \cite{dardenne1993goal, darimont1996formal, van1998managing, van2000handling, van2001goal, letier2004reasoning}, Mylopoulos et al. \cite{mylopoulos1992representing, greenspan1994formal, castro2002towards}, Robinson et al. \cite{robinson2003requirements}, Nuseibeh et al. \cite{nuseibeh1994framework, hunter1998managing}, to name some.

According to the \ZJ{} view, in any concrete systems engineering project, \RE{} is successfully completed when the following conditions are satisfied \cite{zave1997four}:

\begin{quote}
\begin{enumerate}
    \item{``There is a set $R$ of requirements. Each member of $R$ has been validated (checked informally) as acceptable to the customer, and $R$ as a whole has been validated as expressing all the customer's desires with respect to the software development project.}
    \item{There is a set $K$ of statements of domain knowledge. Each member of $K$ has been validated (checked informally) as true of the environment.}
    \item{There is a set SS of specifications. The members of $S$ do not constrain the environment; they are not stated in terms of any unshared actions or state components; and they do not refer to the future.}
    \item{A proof shows that $K, S \vdash R$. This proof ensures that an implementation of $S$ will satisfy the requirements.}
    \item{There is a proof that $S$ and $K$ are consistent. This ensures that the specification is internally consistent and consistent with the environment. Note that the two proofs together imply that $S$, $K$, and $R$ are consistent with each other.''}
\end{enumerate}
\end{quote}

If the satisfaction of these conditions marks the end of \RE{} in any systems engineering project, then we can give a compact formulation of the default problem that \RE{} should solve, which we call the \ZJRPfull{} hereafter.

\zi{\ZJRPfull{}}: Given a set $R$ of requirements, and a set $K$ of domain knowledge, find a specification $S$, such that $S$ satisfies the following conditions:
\begin{enumerate}
    \item{There is a proof of $R$ from $K$ and $S$, written $K, S \vdash R$,}
    \item{$K$ and $S$ are consistent, written $K, S \not\vdash \bot$.}
\end{enumerate}

\subsection{Rationale for Contractual Relationships}\label{s:rationale:contract}
Why are contractual and economic relationships absent in the mainstream account of the requirements problem? 
It is either that contracts and economics do not matter in the \ZJRP, or that there is a lot to say about that problem even without considering the two other dimensions. Neither of these positions is satisfactory. To see why, we need to consider three central, recurrent questions in \RE:
\begin{enumerate}
\item How to decide the relative importance of requirements? This is called the \textit{prioritisation problem} hereafter, and is discussed in Section \ref{s:rationale:contract:prioritisation}. 
\item How to determine if satisfying a requirement is justified, i.e., that we have good enough reasons to believe that \zi{that} requirement should be satisfied? This is called the \textit{acceptability problem} (Section \ref{s:rationale:contract:acceptability}).
\item How to determine if and how well a requirement is satisfied? This is the \textit{validation problem} (Section \ref{s:rationale:contract:validation}).
\end{enumerate}

In the remainder of this section, I will argue that each of these problems is in fact a problem that involves contractual, economic, and engineering relationships. None of them can or should be seen simply as engineering problems.

\subsubsection{Prioritisation}
\label{s:rationale:contract:prioritisation}
The prioritisation problem reflects the inability to simultaneously satisfy all the requirements that we may want to satisfy. The prioritisation problem asks how to decide which ones to satisfy first. 

How much we can prioritise depends on what the requirements are about, how much resources we can commit, and throughput, or roughly speaking, how much of the requirements can be satisfied by how much resources. This, in turn, begs the question of what we expect to gain from that commitment, and therefore, there are economic relationships at play when we do requirements prioritisation \cite{sivzattian2001linking,kukreja2012selecting,achimugu2014systematic,riegel2015systematic}. 

Suppose that you expected the most competent people in the world to work to satisfy your requirements. Would this influence the content of the requirements that you would ask them to satisfy? Would you ask for different requirements if this were not the case, if you in fact knew very little of who would work on them, and what their competence is? 

In \zi{Hertz Corporation v. Accenture LLP} \cite{hertz_v_accenture}, one of many issues that the car rental company raised was that Accenture misled it into believing that ``the best talent in the world'' would work on satisfying the requirements that Hertz had at the time. Paragraph 4 of the Complaint (April 19, 2019) reads as follows:
\begin{quote}
``After Accenture put on an impressive, one-day presentation for the Hertz team that included a demonstration of the transformed Hertz digital experience, Hertz selected Accenture to design, build, test, and deploy Hertz’s new website and mobile applications (or `apps').''
\end{quote}

The Memorandum \& Order (October 25, 2019) adds detail:
\begin{quote}
``Ultimately, Hertz hired Accenture following a one-day marketing presentation, in which Accenture touted its world-class expertise in website and mobile application development. The presentation contained slides stating that Accenture's staff consisted of `800 [e]xperts' who comprised `[t]he best talent in the world.' The presentation also stated `[w]e've got the skills you need to win' and that Accenture would `put the right team on the ground [from] day one.'''
\end{quote}

Hertz claimed that the presentation was misleading, and Accenture's response was that it is ``non-actionable puffery''. This is what the Court concluded as well:
\begin{quote}
``Accenture’s representation that it housed `800 [e]xperts' amounting to `[t]he best talent in the world,' along with its promise that it had `the skills you need to win' and would `put the right team on the ground [on] day one,' are quintessential examples of puffery. Accordingly, this Court concludes that the alleged misstatements in the marketing presentation
are non-actionable.''
\end{quote}

How does this relate to prioritisation? If one prioritises assuming that the best talent would be doing the work, one does it differently than if different talent, so to speak, is expected to do the work: the best talent would presumably do more and better than anyone else.

More importantly, if this was not in a marketing presentation, but was in some more accurate manner defined in the contract, the contract would not only influence what gets prioritised, but also what enters prioritisation in the first place. Going back to \ZJRP, the content of $R$ would be influenced by the contract, and since the contractual relationships are outside of the definition of \ZJRP, it can only be part of a broader problem to solve.

The incentives that a contract defines, or the economic relationships it sets up, also influence how prioritisation will be done. 

Suppose that a contract A pays out the same amount to those satisfying requirements regardless of the expected value of outcomes of different requirements. This would be the case of developing, say, software which streams video online, but getting paid for hours spent making it, regardless of how much it is actually used once up and running. By expected value of outcomes of satisfying a requirement, I mean the value that the system-to-be is expected to generate at run-time if it is in fact designed to satisfy that requirement. 

Suppose that in another contract B, their payout is proportional to actual value at run-time. In the same example, those developing it would be paid relative to advertising revenue, for example, which in turn is proportional to number of times each video is viewed (or some other metric correlating with usage). 

This difference between A and B exists in, for example, in employment contracts for those who accept the obligation to satisfy requirements: case A would be a contract that includes neither shares, nor rights to shares in equity of the legal entity (or in revenue, as in the case of royalties) that commercialises the system-to-be. In case A, there is no reason for them to prioritise or insist to prioritise requirements which may be more difficult to satisfy, but could generate higher benefits, while they may think about this in another way if they can claim some of these benefits, as in case B.

In short, we cannot do requirements prioritisation while ignoring the contract. If you do, as Hertz did, you take on substantial risks.

\subsubsection{Acceptability}
\label{s:rationale:contract:acceptability}
A contract could specify that the party which has the rights to give requirements can indeed give \zi{any} requirements, i.e., give the role of requirement to any proposition. 

If I have that right by that contract, then I can turn any proposition into a requirement, including something as ludicrous as ``green cheese should grow on the Moon''. But even if the contract was written in such a way, and there was someone as delusional as to accept the obligations in it, then I could ignore any complaints they may have. That contract would guarantee \zi{a priori} that all propositions I turn into requirements are acceptable as requirements, regardless of what these propositions are about. This is not realistic. 

One contractual setup that I have experienced in practice in the past is very much the one that seemed to be in place between Hertz and Accenture: Hertz and Accenture had, according to the Court Opinion and Order (March 3, 2020), a ``Consulting Services Agreement'' since 2004, and ``[t]he Project was to be conducted in phases, and the services and deliverables for each phase were, in turn, specified in letters of intent (`LOIs') and corresponding statements of work (`SOWs').'' In such a setup, the SOW would describe those requirements that the party which should satisfy indeed accepts to satisfy. Writing SOWs, in other words, comes after an assessment -- however well or badly done -- of clarity, completeness, and feasibility of producing the solution which is expected to satisfy the propositions suggested to be treated as requirements (or a subset thereof). 

I have also been involved on either side of very different contracts, where the contract does not presuppose the existence of a comprehensive SOW. In place of specifying requirements on the deliverable (or solution to make to satisfy the requirements), it specifies constraints on the process which the parties go through, in order to define requirements after the contract is signed. In place of buying a solution, one buys time from experts who are expected to be able to design and deliver the solution, even if the specifics of that solution, or of the problem to solve may be elusive at contracting time. One of the twelve principles in the Agile Manifesto reads ``[t]he best architectures, requirements, and designs emerge from self-organizing teams'' \cite{beck2001agile}, something that has to be acknowledged through a contract: in the first case above, where a detailed SOW is needed and requirements planned ahead, this principle cannot be satisfied and it is the contract that makes it unsatisfiable. 

The specifics of the acceptability problem that we will face in a given situation can be determined by the contract. For example, the contract may say that any proposition is acceptable as long as it is given by those with the rights to confer the role of requirements to propositions. Another contract may ask that some evidence should be given for the acceptability of requirements, and it would have to define what counts as evidence. A third contract might say that any proposition given up to a specific date can be given the role of requirement, but not past that date. A fourth contract might not specify that date. Consider how different these are, and the extent to which they shape the requirements engineering process that we will have to use to eventually satisfy the requirements. In the third contract, we might find the waterfall process good enough. The fourth contract makes waterfall a more difficult process choice than one inspired by the Agile Manifesto. 

In conclusion, we cannot ignore the contract when picking the requirements engineering process, and specifically how we decide if a proposition can have the role of requirement.

\subsubsection{Validation}
\label{s:rationale:contract:validation}
The validation problem concerns how you determine if a requirement is satisfied, and how well it is satisfied. If the contract specifies how this should be done, then the requirements engineering process cannot ignore it. Paragraph 7 in the Complaint that Hertz filled against Accenture reads as follows:
\begin{quote}
``Hertz relied on Accenture’s claimed expertise in implementing such a digital transformation. Accenture served as the overall project manager. Accenture gathered Hertz’s requirements and then developed a design to implement those requirements. Accenture served as the product owner, and Accenture, not Hertz, decided whether the design met Hertz’s requirements.''
\end{quote}

If the contract specified that Accenture validates the requirements it implements, then that contract determines a key property of the requirements engineering process between these two parties. The point, again, is that we cannot dissociate how one does requirements engineering from the content of a contract which determines who has the right to give propositions the role of requirements.

How does this tie to the \ZJRP, the engineering problem? The issue is who is accountable for the proof that $K, S \vdash R$. At first sight, this should not matter; after all, it is a proof and if it is there, anyone can follow the steps in it and see for themselves. The problem, however, is that proof rests on only those relationships between propositions in $K$, $S$, and $R$ which have been formalised and appear in there. For example, you can have $K = \{ p_1, p_1 \wedge p_2 \rightarrow p_3 \}$, $S = \{ p_2 \}$ and $R = \{ p_3 \}$, and it won't really matter what each of these is about; you will have $K, S \vdash R$. At the same time, $p_1$ may be about green cheese, $p_2$ about pink clouds, and $p_3$ about a property of a car rental company website.

\subsection{Rationale for Economic Relationships}\label{s:rationale:economics}
It may seem from the above that there are two ideas that I am tying to the concept of requirement, one being that there must be a contract, and the other that there must be some economic relationships, some expectations and exchange of value between the parties involved.

These are not separate ideas, and we therefore cannot say that there is an engineering problem, to which I am tying a contracting one, and then a third, economic one - the contract and economics are so closely tied together that we cannot take one and ignore the other. 
Why is this so? The obvious idea is that there is no acceptance of obligation without expectation of value in return (except under duress, which we leave aside). Importantly this observation matches the central tenet of contract law, that contract establishes \zi{chosen} obligation. 
\begin{quote}
``[C]ontracts [...] arise through an exchange of promises. This is inscribed in legal doctrine, in the principles that contracts are created through offer, acceptance, and consideration. An offer, according to the U.S. second Restatement on Contracts, 'is the manifestation of willingness to enter into a bargain, so made as to justify another person in understanding that his assent to that bargain is invited and will conclude it. (R2 Contracts: §24)' '' \cite{sep-contracts-theories}
\end{quote}

For a proposition $p$ to get the role of requirement, we need parties willing to enter the bargain, whereby there will be at least one such party that will expect $p$ to be satisfied through this bargain, and another which expects to get value in exchange for satisfying $p$.

\begin{quote}
``To establish a contract, an offer must be met with an appropriate acceptance, characteristically 'a manifestation of assent to the terms [of the offer] made by the offeree in a manner invited or required by the offer. (§50)'' \cite{sep-contracts-theories}
\end{quote}

For $p$ to be a requirement, then, we need a party which accepts the offer.

\begin{quote}
``These requirements entail that all orthodox contracts contain promises. But not all promises establish contracts, among other reasons because the law further requires that contracts be supported by good consideration. The consideration doctrine, in its modern form, adds a bargain requirement to contract formation. The Restatement says that 

`[t]o constitute consideration, a performance or a return promise must be bargained for'

and adds that

`[a] performance or return promise is bargained for if it is sought by the promisor in exchange for his promise and is given by the promisee in exchange for that promise. (R2 Contracts: §71)'

Contracts, that is, must arise not out of a simple, gratuitous promise, but rather out of an exchange of promises.'' \cite{sep-contracts-theories}
\end{quote}

Finally, then, we do not have promises going in one direction only. Those asking for $p$ to be satisfied need to promise, at the very least, that value will be given to those who promise to satisfy $p$, i.e., to perceive that satisfying $p$ is a requirement for them.

But don't we have promises when, for example, I ask you to give me your book, offer you \$20 in exchange, and you accept? Am I giving a requirement? It is an exchange, but does it require a contract? 

\begin{quote}
``[T]he main economic function of contract law is to assist transacting parties who face difficulties associated with \zi{non-simultaneous transactions}. Stated differently, contract law facilitates \zi{deferred exchanges}. An example is useful here. Suppose that you agree to build me a boat in exchange for £10,000, payable in advance. Absent a law of contract, there is an obvious risk that, having received the £10,000, you will renege on our deal and pocket the money.'' \cite{smith2004contract}
\end{quote}

When we talk about requirements which involve making something -- and we say a new system or a system-to-be typically in requirements engineering -- then this is a deferred exchange, and it makes little sense therefore to ignore ensuring a contractual framework and being clear about economic relationships.

In conclusion, contract, economics and requirements are inseparable.

\section{Solution}\label{s:solution}
The necessary and sufficient conditions for a proposition $p$ to have been given the role of a requirement are as follows.
\begin{enumerate}
    \item There is a so-called \zi{\RC}, which defines
    \begin{enumerate}
        \item the right to give propositions the role of requirements, called the \textit{right to request} hereafter,
        \item the \textit{obligation to satisfy requirements},
        \item the \textit{obligation to validate} if a product satisfies requirements,
        \item the \textit{obligation to remunerate satisfaction} of requirements,
        \item the \textit{obligation to remunerate validation},
        \item the \textit{right to request remuneration for satisfying requirements},
        \item the \textit{right to request remuneration for validating requirements}.
    \end{enumerate}
    \item The \RC{} is enacted, i.e., there is a party for each role that the contract creates by defining the respective rights and obligations.
    \item Rights and obligations in the \RC{} are, respectively exercised and discharged, and specifically, the party which holds the right to request indeed requests that $p$ be satisfied, and thereby, $p$ gets the role of a requirement.
\end{enumerate}

\subsection{Departure}\label{s:solution:departure}
This approach to defining how a proposition gets to be called a requirement is different from related work mentioned so far in this paper. There is no need to be concerned with grammatical mood, speech acts, or where the proposition appears in the refinement tree. The concept of requirement is separated from intentional states and folk psychology; this addresses the problem I raised elsewhere \cite{DBLP:conf/er/Jureta17}: intentional states cannot be known (i.e., while you may experience or observe your intentional states, you cannot do so with mine, nor can I do so with yours), so any appeal to intentional states as reasons for a proposition to be a requirement essentially asks you to assume that there is something behind the requirements that you have no way of observing, or otherwise accessing. It puts a veil of mystery where there is no need for one. Instead, the explanation for a proposition to have the role of requirement is that there is an enacted contract and parties exercising their rights and discharging their obligations.

The solution above emphasises the contract in the first condition. That first condition is not enough, however. 

The contract must be enacted, the rights and obligations in it should be, respectively, accepted and discharged. This is captured in the second and third conditions. This is also where economic relationships come into play: if there are no expectations of value from making the various promises formalised in the contract, then there will be no one to enact it.

\subsection{Network}\label{s:solution:network}
The necessary and sufficient conditions given earlier can be represented as a network of relationships over expectations, rights, obligations, actions, and outcomes. That representation is called the \textit{Requirements Contract Network} (also only ``Network'' hereafter). I use the Network to discuss alignment of interests of the parties in the \RC, and how the contract can be designed to support that alignment.

The Network is shown in Figure \ref{fig:f5}. Each node is an expectation, action, or outcome. The \RC{} itself appears through black nodes, which represent actions involving rights and obligations that the contract defines. The network is arranged in the Figure to show sequence over time, with time passing from left to right. 

\subsection{Nodes and Links}
In the Network, everything starts from the expectations, which lead to actions related to the \RC, namely the acceptance of rights and obligations, which are, respectively, exercised and discharged through subsequent actions. Outcomes result from actions.

Each link reads ''is necessary for'', a relationship indicating that the target of the link cannot happen if the source of the link hasn't. Taken together, all links targeting a specific node are the set of sufficient conditions for what that node describes: expectations to be had, actions to be executed, or outcomes to occur. 

\subsection{Roles}
\begin{figure}[t]
\centering
\includegraphics[width=\textwidth]{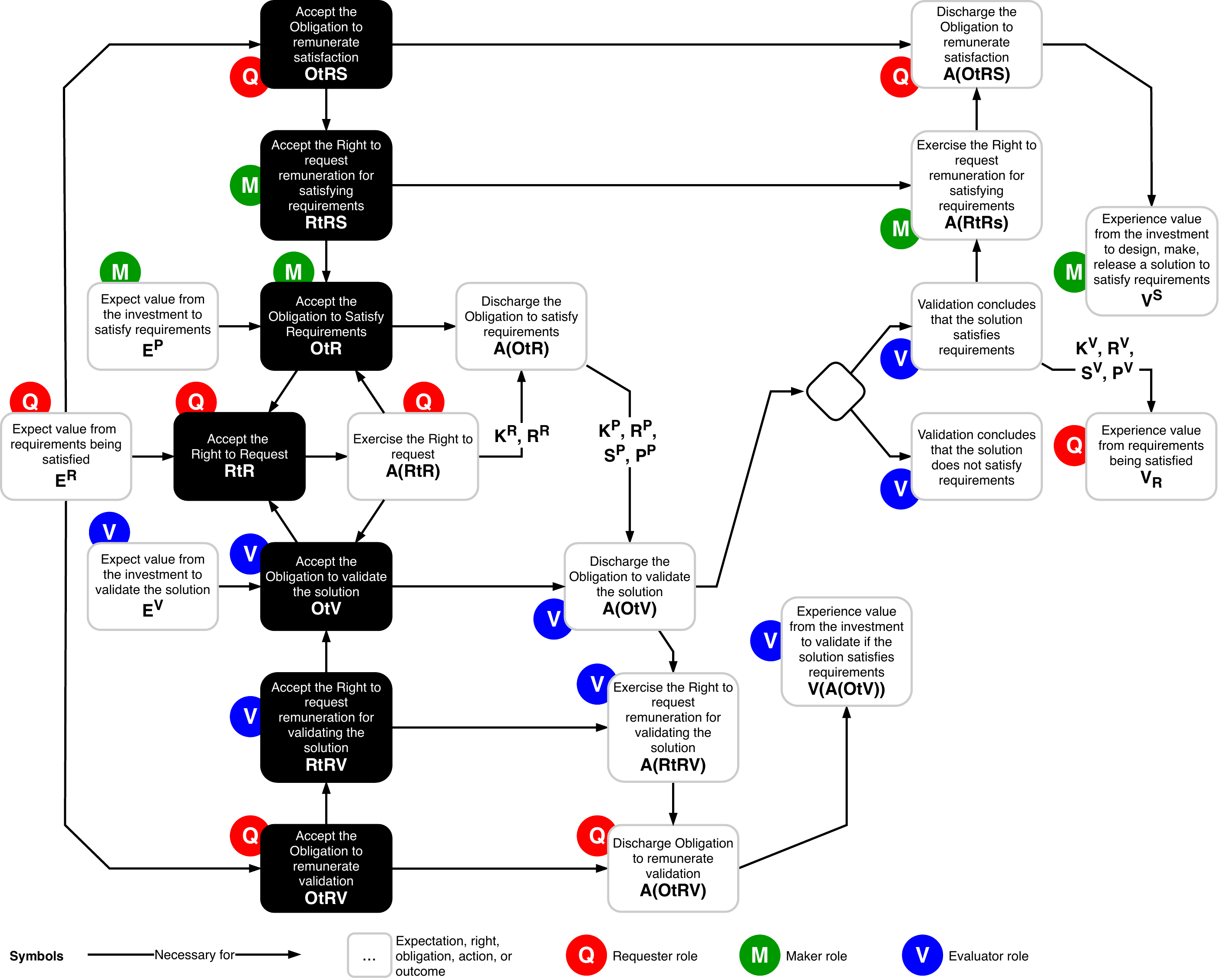}
\label{fig:f5}
\caption{Nexus of rights, obligations, actions, expectations, and outcomes subsumed by a \RC}
\end{figure}

Each expectation, action, and outcome is associated with a role. Roles are placeholders for parties who will fill them when an actual \RC{} is enacted. Moreover, roles shown in Figure \ref{fig:f5} are merely one way in which expectations, actions, and outcomes can be grouped together and assigned to parties. Three roles are shown, and should be read as follows.
\begin{itemize}
\item Requester role, labelled \wx{Q} in the Figure, is to be filled by the party which has the expectation of getting value if her requirements are satisfied.
\item Maker role, labelled \wx{M}, is responsible for making the product that should satisfy requirements; the party in this role expects value from doing so.
\item Evaluator role, labelled \wx{V}, is responsible for evaluating if the product satisfies requirements, and the party in this role expects value from doing the evaluation. 
\end{itemize}

\subsection{Roles and Parties}
It is also not necessary that each role is filled by a different party. For example, in both \zi{Hertz Corporation v. Accenture LLP} and \zi{GB Gas Holdings Ltd v Accenture (UK) Ltd \& Ors} \cite{centrica_v_accenture}, Accenture filled itself both the Maker and Evaluator roles. Whether that is the best choice is a separate question; there is a lot to say about how many parties you may want to involve in a \RC, depending on which role you have; I return to this in Section \ref{s:alignment}.

\subsection{Right to Request}
The starting point is the expectation of value from having requirements satisfied (denoted $\wx{E}^\wx{R}$ in the Figure). There is no reason for the \RC{} to exist if there is no such expectation. In order for a party to accept the right to request, \wx{RtR}, and provide requirements that need to be satisfied, three conditions need to hold: 
\begin{itemize}
\item The party which is to accept the right to request has the expectation of value from seeing those requirements satisfied ($\wx{E}^\wx{R}$),
\item A party accepts the obligation to satisfy requirements (\wx{OtR}),
\item A party accepts the obligation to validate if requirements are satisfied (\wx{OrV}).
\end{itemize}

\subsection{Obligation to Satisfy}
No one will accept the obligation to satisfy requirements (\wx{OtR}) if they lack an expectation of value from doing so ($\wx{E}^\wx{P}$) and an idea of what the requirements and assumptions around them may be; the latter comes from having at least some requirements and assumptions communicated by the party which accepted the right to request (the party having the right to request should also exercise that right). 

The interplay between knowing requirements, and accepting the obligation to satisfy requirements, is captured in two ways. The first is the need for the right to request to be exercised, in order to accept the obligation to satisfy requirements (link from \wx{A(RtR)} to \wx{OtR}). The second is the overlap of acceptance of the right to request with the acceptance of the obligation to satisfy requirements, and then the overlap of the latter with the exercising of the right to request. 

\subsection{Obligation to Validate}
Same applies to validation, in that one needs to expect value from doing it ($\wx{E}^\wx{V}$) and know at least some of what they are getting themselves into (the right to request needs to be exercised), in order to accept the corresponding obligation (\wx{OtV}). 

In addition, to accept the obligation to satisfy requirements \wx{OtR}), the party needs to have the right to request remuneration for the investment that she will have to make to actually do so (\wx{RtRS}), i.e., needs a way to request value. At the same time, it will make sense for her to accept that right to ask for remuneration if someone else accepts the obligation to provide it (\wx{OtRS}). Note how this creates a nexus that links expectations to the obligations and rights in the \RC.

We have the analogous situation for validation. For someone to accept the obligation to validate if requirements are satisfied (\wx{OtV}), she needs to accept the right to request remuneration (\wx{RtRV}), and thereby be able to request value she expects. In turn, a party needs to accept the obligation to provide that remuneration for having validated if requirements are satisfied (\wx{OtRV}). 

All relationships identified above mean that expectations need to be there, as well as rights and obligations accepted, before the Right to request is exercised, and requirements are given. 

\subsection{Imperfect Transfer}
The party which holds the right to request will exercise that right and provide two sets of propositions:
\begin{itemize}
\item A set $\wx{K}^\wx{R}$ of propositions that convey this party's understanding of her current situation, one which gave rise to her requirements, and
\item A set $\wx{R}^\wx{R}$ of propositions that, by exercising the Right to request, this party gives as requirements.
\end{itemize}

As different parties provide and satisfy requirements, communication between them means that there is a difference between one's assumptions and requirements and other's understanding of these assumptions and requirements. This may be, e.g., because of incompleteness, vagueness, ambiguity, or other deficiencies which come from the impossibility to be perfectly clear and comprehensive in providing all the assumptions and all the requirements that one may want to provide, or considers implied in that which one does in fact provide explicitly. It goes back to the distinction between explicit and implicit knowledge \cite{dienes1999theory}. 

The implication here, is that $\wx{K}^\wx{R}$ and $\wx{R}^\wx{R}$ are one party's, while the other will work on something more or less different. Specifically, the party which discharges the obligation to satisfy requirements, will have the assumptions $\wx{K}^\wx{P}$, have understood requirements $\wx{R}^\wx{P}$. As the output of this party's work, we have the specification of the product intended to satisfy requirements once in use, $\wx{S}^\wx{P}$, and the product itself $\wx{P}^\wx{P}$. (The difference between $\wx{S}^\wx{P}$ and $\wx{P}^\wx{P}$ is that the former is the blueprint of the thing, and the latter the thing itself.)

$\wx{K}^\wx{P}, \wx{R}^\wx{P}, \wx{S}^\wx{P}, \wx{P}^\wx{P}$ are the outputs of the investment to satisfy the requirements. 

In order to discharge the obligation to validate the product (\wx{A(OtV)}), the party who accepted to do this, will do it on the basis of its own interpretation of these outputs, namely $\wx{K}^\wx{V}, \wx{R}^\wx{V}, \wx{S}^\wx{V}, \wx{P}^\wx{V}$. Once it discharges this obligation, that party will exercise its right to request remuneration (\wx{A(RrRV)}), which leads to it to experience value (denoted \wx{V(A(OtV)}) in Figure \ref{fig:f5}).

The result of validation will be either that the product does satisfy requirements, or that it does not. If it does, the party who expected value from having requirements satisfied will experience value (denoted $\wx{V(P}^\wx{R}\wx{)}$). In addition, the party who expected value from satisfying requirements will experience some value (denoted \wx{V(A(OtR)}). 

\subsection{Handling Failure}
If validation leads to the conclusion that the product does not satisfy requirements, we need to return to the \RC, and the rules it specifies around the handling of exceptions, including failure of this kind. I leave these out of scope in this paper; one option is that failure to satisfy requirements leads to going back to exercising the right to request (\wx{A(RtR)}).

\section{Alignment}\label{s:alignment}
What are the interests of each party that decides to accept its rights and obligations in the \RC? Are these interests aligned? When are they aligned? What if one is pursuing actions which improve its outcomes, but in doing so do not improve those of others? Why would they not be aligned? Is it possible to design the \RC{} in ways that improve alignment? 

Alignment is discussed in three steps in this Section. Firstly, we need to make assumptions about the interests that each party in the \RC{} may have. From those assumptions, we can catalogue sources of misalignment, which can be shown using the Network in Figure \ref{fig:f5}. Finally, we can discuss how the \RC{} can be designed to reduce the likelihood of specific types of misalignment to occur.

\subsection{Interests}\label{s:alignment:interests}
Why would a party accept rights and obligations in a \RC? According to the Network in Figure \ref{fig:f5}, the answer is as follows.
\begin{itemize}
\item There is a party which expects value if requirements are satisfied, $\wx{E}^\wx{R}$, and is willing to invest to get this value -- both satisfaction of requirements, and evaluation (validation) of the product need to be remunerated.
\item There is a party which expects value $\wx{E}^\wx{P}$ because it invests its own resources to satisfy requirements by designing, making, and delivering a product to the party which had requirements in the first place.
\item There is a party which expects value $\wx{E}^\wx{V}$ because it invests its own resources in evaluating if the product satisfies requirements.
\end{itemize}
To simplify writing, I will refer to these parties through the roles they would be filling in the Network in Figure \ref{fig:f5}: Requester expects $\wx{E}^\wx{R}$, Maker expects $\wx{E}^\wx{P}$, and Evaluator expects $\wx{E}^\wx{V}$.

What does expected value depend on? Each party needs to make an investment in order to produce that which eventually yields value for them. Thus, at the very least, expected value will be a function of expected benefits and expected costs; let's write these as follows.
\begin{align}
    \wx{E}^\wx{R} &= E(\wx{B}^\wx{R}) - E(\wx{C}^\wx{R}) \\
    \wx{E}^\wx{P} &= E(\wx{B}^\wx{P}) - E(\wx{C}^\wx{P}) \\
    \wx{E}^\wx{V} &= E(\wx{B}^\wx{V}) - E(\wx{C}^\wx{V})
\end{align}
Since the Requester accepts the obligation to remunerate the satisfaction of requirements (the node OtRS is labeled Q in Figure \ref{fig:f5}), the benefits that Maker can expect are capped by the cost that the Requester is willing to bear. At the same time, the Requester remunerates evaluation, so that the expected benefits of the Evaluator are also capped by the Requester's expected cost. More specifically, we have
\begin{align}\label{eq-146879}
    E(\wx{B}^\wx{P}) + E(\wx{B}^\wx{V}) \leq E(\wx{C}^\wx{R})
\end{align}
It has been implicit so far, but it is important to note now that expected value for each of these roles should be positive, else there is no apparent reason for the losing party to enter the \RC. 

\begin{assumption}
For a party to consider entering into an \RC, its expected benefits should outweigh its expected costs.
\end{assumption}

\begin{align}
    E(\wx{B}^\wx{R}) &> E(\wx{C}^\wx{R}) \\
    E(\wx{B}^\wx{P}) &> E(\wx{C}^\wx{P}) \\
    E(\wx{B}^\wx{V}) &> E(\wx{C}^\wx{V})
\end{align}
Does this mean that a party should enter an \RC{} as soon as expected benefits outweigh expected costs? Is that the aim that each party has, when entering the contract? If, as usually in mainstream economics \cite{}, expected value is expected utility, then one enters the contract in the aim of maximising one's utility. In other words, if we have three roles in the contract, and we assumed each expects value from exercising its rights and discharging its obligations in the contract, then we also have three parties (if each role is occupied by a different party). 

\begin{assumption}
Each party in the contract will make decisions which it perceives as maximising the value that she will actually receive after exercising the rights and discharging the obligations that she accepted by accepting the \RC.
\end{assumption}

The objective functions of the parties, provided that there are three of them, are as follows.
\begin{align}
    \text{Requester:}\ & \textit{max}\ \wx{V}^\wx{R} \\
    \text{Maker:}\ & \textit{max}\ \wx{V}^\wx{P} \\
    \text{Evaluator:}\ & \textit{max}\ \wx{V}^\wx{V}
\end{align}
I will also assume that expected and actual value are almost the same. I will challenge this assumption later.

\begin{assumption}
Each party in the \RC{} will exercise its rights and discharge its obligations in such a way that makes actual value as close as possible to its expected value.
\end{assumption}

\begin{align}
    \wx{V}^\wx{R} \approx \wx{E}^\wx{R} \\
    \wx{V}^\wx{P} \approx \wx{E}^\wx{P} \\
    \wx{V}^\wx{V} \approx \wx{E}^\wx{V}
\end{align}

To keep the discussion simple still, I need to make another brittle assumption, to relate maximisation of actual value and of expected value. I will question this assumption later too.

\begin{assumption}
At the time when a party considers entering into a \RC, i.e., accepting a role in it and the accompanying rights and obligations, that party will maximise its expected value in order to maximise its actual, future value.
\end{assumption}

\begin{align}
    \text{Requester:}\ & \textit{max}\ \wx{V}^\wx{R} \equiv \textit{max}\ \wx{E}^\wx{R} \\
    \text{Maker:}\ & \textit{max}\ \wx{V}^\wx{P} \equiv \textit{max}\ \wx{E}^\wx{V} \\
    \text{Evaluator:}\ & \textit{max}\ \wx{V}^\wx{V} \equiv \textit{max}\ \wx{E}^\wx{V}
\end{align}

\subsection{Simple Case of Conflict of Interest}
Consider now four situations that a party can be in, and for simplicity, let that party be the Requester.
\begin{itemize}
\item The party believes that each incremental unit of expected cost adds more than that in expected benefits: there is a gain to be made by taking on more expected costs, that is, the following is true:
\begin{equation}
    \frac{\Delta E(\wx{B}^\wx{R})}{\Delta E(\wx{C}^\wx{R})} > 1
\end{equation}
If so, then maximising expected value means deciding to invest more, up to some limit which is private information for that party. 
\item The party believes that each incremental unit of expected cost adds less than that in expected benefits: it costs disproportionately more to get an increment in expected benefits:
\begin{equation}\label{eq-472872}
    0 < \frac{\Delta E(\wx{B}^\wx{R})}{\Delta E(\wx{C}^\wx{R})} < 1
\end{equation}
If this party has a threshold of expected benefits, she should reduce costs up to -- if that happens at all -- the point when she believes that marginal expected value is equal to marginal expected cost, that is, up to the point when:
\begin{equation}\label{eq-1124}
    \frac{\Delta E(\wx{B}^\wx{R})}{\Delta E(\wx{C}^\wx{R})} = 1
\end{equation}
\item The party believes that any additional unit of expected cost leads to an equal unit of expected benefits, as in Equation \ref{eq-1124}. If she hasn't reached a private maximal expected cost that she is willing to bear, her interest is to invest more, i.e., to increase her expected cost as long as marginal expected cost equals marginal expected benefit, i.e., until Equation \ref{eq-472872} is true, or until she has reached the maximal expected cost she is willing to accept. 
\item Finally, we can have this situation:
\begin{equation}\label{eq-472873}
    \frac{\Delta E(\wx{B}^\wx{R})}{\Delta E(\wx{C}^\wx{R})} < 0
\end{equation}
If this is because $\Delta E(\wx{B}^\wx{R}) < 0$, then any increase in expected cost is believed by that party to lead to a reduction in expected benefits. If the fraction above is negative because $\Delta E(\wx{C}^\wx{R}) < 0$, then any reduction in expected cost is believed to increase expected benefit.
\end{itemize}

But this is only one party, yet we have three in the \RC. What happens if we have the following situation?
\begin{equation}
    \frac{\Delta E(\wx{B}^\wx{R})}{\Delta E(\wx{C}^\wx{R})} < 1, \frac{\Delta E(\wx{B}^\wx{P})}{\Delta E(\wx{C}^\wx{P})} > 1, \text{and } \frac{\Delta E(\wx{B}^\wx{V})}{\Delta E(\wx{C}^\wx{V})} > 1
\end{equation}
Due to Equation \ref{eq-146879}, we can rewrite this as follows:
\begin{equation}
    \frac{\Delta E(\wx{B}^\wx{R})}{\Delta (E(\wx{B}^\wx{P}) + E(\wx{B}^\wx{V}))} < 1, \frac{\Delta E(\wx{B}^\wx{P})}{\Delta E(\wx{C}^\wx{P})} > 1, \text{and } \frac{\Delta E(\wx{B}^\wx{V})}{\Delta E(\wx{C}^\wx{V})} > 1
\end{equation}
The above shows a conflict of interest: interests of the Maker and Evaluator are to increase costs in order to increase their benefits more, while this isn't in the interest of the Requester. 

This only scratches the surface of interest alignment. There are more than four situations situations identified above, that a party can be in. 

\subsection{Interest Cases}
\begin{figure}[t]
\centering
\includegraphics[width=\textwidth]{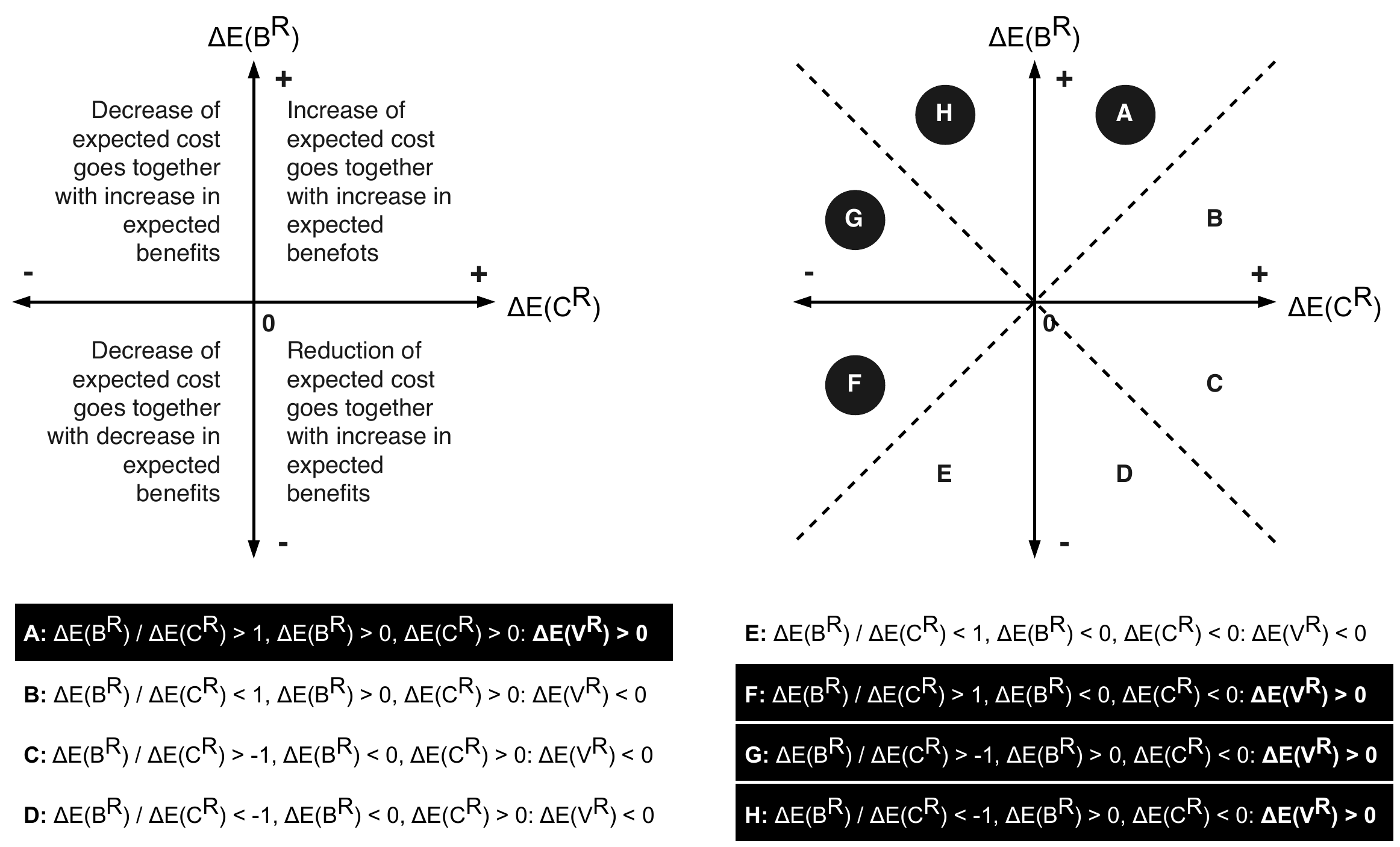}
\label{fig:cl1}
\caption{Interest Cases of the Requester}
\end{figure}

Figure \ref{fig:cl1} shows eight cases, labelled clockwise from A to H. Each is called an interest case, in that it describes what a party should have as its interest, assuming the given relationship between the change of benefits and change of costs. 

Figure \ref{fig:cl1} has three parts. The upper left corner shows four quadrants made by  positive and negative change of expected benefits and costs of the Requester, and the reading of points in each quadrant. The upper right corner of the Figure shows the same quadrants, now split into eight combinations of positive and negative changes in expectations of benefits and costs. The lower part of the Figure shows in detail all eight combinations, and highlights those where a change in expected benefits and costs leads to an increase of expected value for the Requester. It shows that the Requester should want to be in one of the four situations A, F, G, or H, rather than others.  

In summary, Figure \ref{fig:cl1} shows what the interest of the Requester should be, depending on the interest case it is in. It is important to note that this same set of interest cases applies to any party; For example, for the Maker role, the same interest cases apply, except that you need to replace $\wx{B}^\wx{R}$ with $\wx{B}^\wx{P}$ and $\wx{C}^\wx{R}$ with $\wx{C}^\wx{P}$.

In A in Figure \ref{fig:cl1}, positive increase expected cost goes together with a positive increase in benefits, and the increase in the latter is higher than the increase in the former, so there is an interest for the Requester to increase expected costs, since doing so increases expected benefits more: increasing cost will increase expected value.

In F, if the Requester makes a change, that change will involve both a reduction in expected cost and in expected benefits, with expected benefits decreasing slower than the expected costs, hence increasing expected value. 

In G, decrease of expected costs comes with increasing expected benefits, and so, an increase in expected value. 

In H, the dynamics are the same as in G, except that expected benefits increase faster.

\section{Conclusions and Open Questions}
The intent behind this paper is to stimulate a richer discussion of how propositions get to have the role of requirements. If you adopt the perspective that is offered here, that requirements exist in a nexus of economic, contractual, and engineering relationships, new questions come up. What properties do typical requirements contracts actually have (such as the one between Hertz and Accenture)? What are the optimal properties of a requirements contract? How do we design the requirements contract to help align interests of the parties involved? 

\bibliographystyle{plain}
\bibliography{refs}

\end{document}